# A Multifractal Dynamical Model of Human Gait

Bruce J. West[1,2] and Nicola Scafetta[2]

1) Mathematics Division, Army Research Office, Research Triangle Park, NC 27709
2) Physics Department, Duke University, Durham, NC 27705

**Summary:** Walking is regulated through the motorcontrol system (MCS). The MCS consists of a network of neurons from the central nervous system (CNS) and the intraspinal nervous system (INS), which is capable of producing a syncopated output. The coupling of the latter two systems produces a complex stride interval time series that is characterized by fractal and multifractal properties that depend upon several biological and stress constraints. It has been shown that: (i) the gait phenomenon is essentially a rhythmic cycle that obeys particular phase symmetries in the synchronized movement of the limbs; (ii) the fractal and multifractal nature of the stride interval fluctuations become slightly more pronounced under faster or slower paced frequencies relative to the normal paced frequency of a subject; (iii) the randomness of the fluctuations increases if subjects are asked to synchronize their gait with the frequency of a metronome or if the subjects are elderly or suffering from neurodegenerative disease. Here we present a new model, called the super central pattern generator, able to reproduce these known properties of walking and discuss the physiological and psychological interpretations of the model parameters.

## 1. Introduction

The past decade or so has witnessed the development of an area of medicine that can best be described as fractal physiology [1-3]. It has been known for a long time that biophysical time series are stochastic, but it is only more recently that these time series have been identified as fractal and as being generated by scaling phenomena. An example of such time series consists of the beat-to-beat intervals of the human heart, called the heart rate variability (HRV) time series. Peng et al. [4] were the first to show that by a judicious processing of the time series the scaling of the central moments yield the fractal dimension of the cardiovascular control system. It was shown in a number of subsequent studies [5] that the HRV time series, rather than being monofractal, are in fact multifractal. Multifractality means that the fractal dimension of the cardiovascular control system is not constant in time, but changes from instant to instant to more efficiently respond to anticipated changes in the environment. Walking is another phenomenon that is described by scaling time series when looked at properly. Hausdorff et al. [6] were the first to show that the stride-to-stride interval time series, called stride rate variability (SRV), manifest scaling in a way similar to HRV time series. The SRV time series were subsequently shown to also be multifractal rather than monofractal [7]. It is this latter behavior that is of concern to us and today we present a nonlinear dynamical model [8] able to reproduce the known scaling properties of the SRV time series. The new model has a number of parameters for which we provide physiological and psychological interpretations.

Walking is a complex process that we have only recently begun to understand through the application of nonlinear data processing techniques to stride interval time series [6-13]. Walking consists of a sequence of steps partitioned into two phases: a stance phase and a swing phase. The stance phase is initiated when a foot strikes the ground and ends when it is lifted. The swing phase is initiated when the foot is lifted and ends when it strikes the ground again. The stride interval is the time to complete each cycle. Typical stride interval time series [14] for an adult subject in both free and metronome constrained walking at normal, fast and slow regimes are shown in Fig. 1.

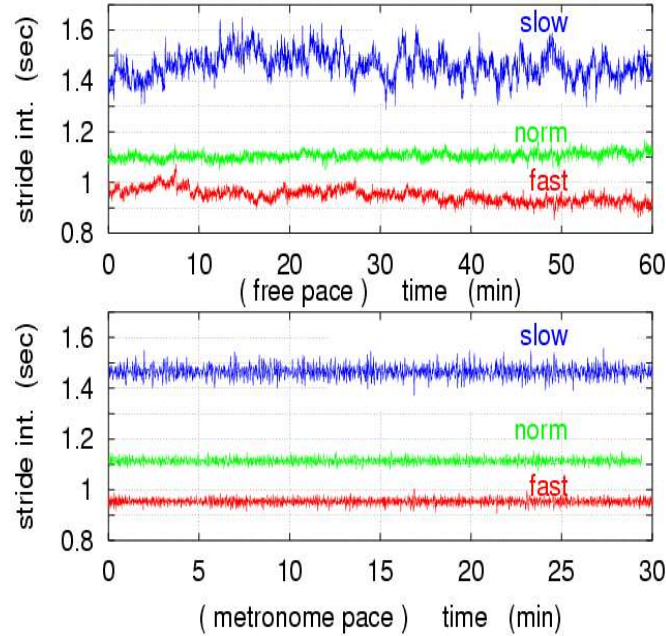

**Figure 1**: Typical stride interval time series in the free and metronome-paced conditions for normal, slow and fast paces (data taken from PhysioNet [14]).

It is evident from Figure 1 that the stride interval of humans fluctuates in time. In fact, it has been known for over a century that there is a variation in the stride interval of humans during walking of approximately 3-4%. Moreover, the random variability of the gait has been shown [6,7] to exhibit long-time correlations, and suggests that the phenomenon of walking is a self-similar, fractal, activity. The fractal and multifractal properties of the stride interval time series were studied using the distribution of the local Hölder exponents via wavelet transforms [15,16], as shown recently by Scafetta et al. [7]. The data were collected and analyzed by Hausdorff et al. [9,11] to determine the dependence of the fractal dimension of the time series on changes of the average rate of walking. These data contain the stride interval time series for ten healthy young men walking at a slow, normal and fast pace, for a period of one hour. The same individuals, at a later time, were requested to walk at a pace determined by a metronome set at the average slow, normal and fast paces for thirty minutes to generate a second data set. Other data were collected from five healthy elderly subjects as they walked for fifteen minutes each, and from five subjects with Parkinson's disease as they walked for six minutes [14] each. All of these time series are multifractal, as we show below.

## 2. Methods

An example of the histograms of Hölder exponents for experimental data is shown in Fig 2. The histograms are fit with Gaussian functions centered on $h_0$ and with a standard deviation σ. We recall that the Hölder exponent $h$ is related to the Hurst exponent [17] $H$ for a monofractal noise by $h=H-1$ for an infinitely long time series. According to this definition, the autocorrelation function [18] of fractal noise is related to the Hurst exponent $H$ and to the Hölder exponent $h$ as follows

$$C(r) \propto r^{-2H-2} = r^{-2h} \qquad (1)$$

where $r$ is the distance between two points in the sequence. An equivalent representation is provided by the power spectrum, the Fourier transform of the autocorrelation function,

$$S(f) \propto f^{1-2H} = f^{-1-2h}, \qquad (2)$$

and $f$ is the frequency. Consequently, $h = 0$ corresponds to pink or $1/f$-noise; and a Hölder exponent in the interval $-1 < h < -0.5$ can be interpreted as antipersistent noise; $h = -0.5$ corresponds to uncorrelated Gaussian noise; $-0.5 < h < 0$ corresponds to persistent noise; $h = 0.5$ corresponds to Brownian motion and $h = 1$ corresponds to black noise. As explained in Ref [7], the fractal properties may be approximately estimated through the mean value $h_0$ of the Gaussian fitting functions, and the multifractal properties are estimated through the ratio $S = \sigma/\sigma_F$, where $\sigma_F$ is the width of a monofractal computer-generated sequence of the same length as the data sequence and having the same Hurst exponent $H=1+h_0$ as the data. A ratio value $S \approx 1$ indicates monofractality, whereas a value $S >1$ indicates multifractality.

### 1.1 Results

By estimating the distribution of Hölder exponents, it has been shown [7] that the stride interval time series for normal gait shows fractal properties similar to $1/f$-noise and is weakly multifractal. The time series may be non-stationary and its fractal variability changes in different gait mode regimes. In particular, the persistence, as well as the multifractality of the stride interval time series, tends to increase slightly for both slow and fast paces, above that of the normal pace. By averaging the results for 10 subjects [6,13], we obtain $h_{0,n}$ = -0.09±0.04, $h_{0,s}$ = 0.035±0.09, $h_{0,f}$ = -0.045±0.06 and the corresponding ratios $S_n$ =1.04±0.04, $S_s$ =1.08±0.10 and $S_f$ =1.05±0.04 for free normal, slow and fast paces, respectively [7]. If the pace is constrained by a metronome, beating at the average rate of the cohort of walkers, the stochastic properties of the stride interval time series change significantly in a wide range, spanning both persistent and antipersistent stochastic processes. In general, in each case there is a reduction in the long-term memory and an increase in randomness as the shift of the Hölder exponent histogram in Fig. 2 shows. By averaging the results for 10 subjects [7,13], we obtain for the average Hölder exponents, in the metronome-constrained case, $h_{0,n}$ = -0.26±0.17, $h_{0,s}$ = -0.48±0.30, $h_{0,f}$ = -0.36±0.23 and for the ratio of standard deviations $S_n$ =1.04±0.07, $S_s$ =1.06±0.07 and $S_f$ =1.09±0.09 for metronome normal, slow and fast paces, respectively [7]. By repeating the Hölder exponent distribution analysis for 5 elderly subjects [13] and a different 5 subjects with Parkinson's disease [14], we find on average $h_{0,eld}$= -0.28±0.10, $h_{0,pd}$= -0.23±0.19 and $S_{eld}$=0.95±0.13, $S_{pd}$= 1.03±0.15. Consequently, for the elderly or those with neurodegenerative diseases the Hölder exponents of the stride interval time series decrease on average from the values for young healthy individuals. The Hölder exponents are all close to the fractal value *h=-0.25* and the multifractality has substantially more variability from person to person than for healthy youth.

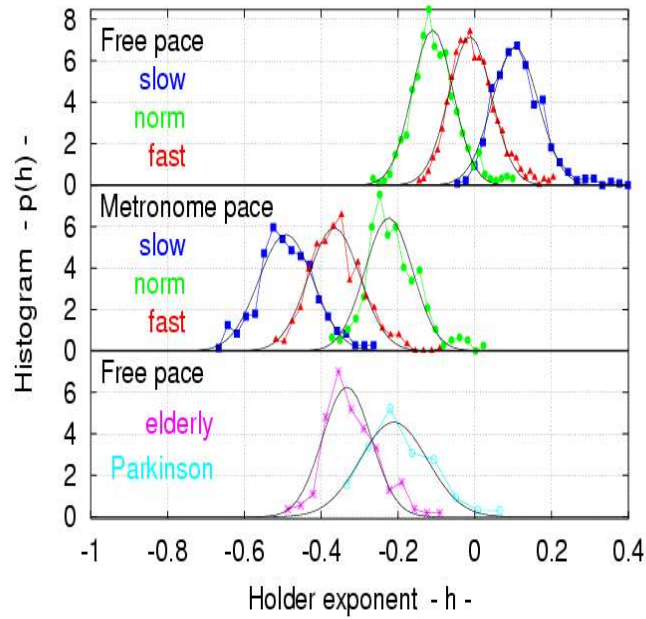

**Figure 2**: Typical Hölder exponent histograms for the stride interval series in the freely walking and metronome conditions for normal, slow and fast paces and for elderly and for a subject with Parkinson's disease. In the data are reported using the average over the subjects. The average histograms are fitted with Gaussian functions.

## 3. The Model

Even though walking is a voluntary action, the typical patterns shown by the stride interval time series suggest particular neural mechanisms that can be mathematically modeled. Traditionally the legged locomotion of animals is understood through the use of an intraspinal network of neurons capable of producing a syncopated output [19-23]. Collins and Richmond [19] modeled this intraspinal network as a hard-wired central pattern generator (CPG) with the coupling of forced nonlinear oscillators, for example, the van der Pol oscillator. These nonlinear oscillators are able to produce limit cycles, simulating the stride interval and when coupled together are sufficiently robust to mimic the phase symmetries of the locomotion observed in the movements in quadrupeds [19-

21] such as the switching among the multiple gait patterns of walking, trotting, cantering and galloping.

The fractal nature of the stride interval time series for humans was incorporated into a dynamic model by Hausdorff et al.[9], using a stochastic version of a CPG. This stochastic model was later extended by Ashkenazy et al. [11,24] so as to describe the changing of gait dynamics during maturation, that is, as humans develop from childhood to adulthood. The model is essentially a random walk on a Markov or short-range correlated chain, where each node of the chain is a neural center that fires an action potential with a particular intensity when visited by the random walker. This mechanism is found to generate a fractal process, with a multifractal width that depends parametrically on the range ρ of the random walker's step size.

### 3.1 Super Central Pattern Generator

The hard-wired CPG and the stochastic CPG capture complementary aspects of the gait phenomenon. Herein we propose that the gait phenomenon be described by a super CPG [8] which is a combination of the above two models. The SCPG is intended to describe the activity of the entire MCS. The SCPG consists of a stochastic CPG that simulates the correlated firing activity of the CNS, coupled to a hard-wired CPG simulating the INS that generates the rhythm of gait. In the SCPG two parameters, the mean gait frequency $f_0$ and the intensity $A$ of the forcing component of the nonlinear oscillator, are sufficient to determine both the fractal and multifractal variability of human gait under normal, stressed and metronome-stressed conditions. Moreover, the decrease of a third parameter, the correlation length, $r_{0,n}$, in the normal pace regime, measures the correlation between the nodes of the CNS, and is sufficient to recover the change of the fractal properties of the stride interval in the elderly and those suffering from neurodegenerative diseases. Finally, even if here we focus on the biped gait, the switching among multiple gait patterns like walking, trotting, cantering and galloping of the quadrupeds are expected to be recovered with the appropriate coupling of several nonlinear oscillators as done by Collins et al [19,20].

The physiological interpretation of the SCPG is that the CNS fires a correlated sequence of action potentials that activate the INS initiating gait. Experiments with cats having spinal cord transactions show that the rate of stride is determined by the intensity of the electrical stimulation, not by its frequency [25]. Consequently the more intense the action potential, the faster the legs move. In this way the CNS is assumed to induce only a sequence of virtual initial frequencies of the gait. The stride interval in gait is given by the time interval that the INS needs to conclude an actual cycle. At the conclusion of each cycle a new pulse is fired by the CNS and a new cycle is initiated. This mechanism is modeled assuming that a stochastic CPG modeling of the CNS produces a correlated sequences of pulses $\{X_j\}$ that generate a sequence of inner virtual frequencies $\{f_j\}$. These virtual frequencies are progressively used in a forced van der Pol oscillator that simulates the cycle of the INS

$$\ddot{Z} + \mu(Z^2 - p^2)\dot{Z} + (2\pi f_j)^2 Z = A\sin[2\pi f_0 t] \tag{3}$$

where over-dots denote derivatives with respect to time. The stride interval is the actual period of each cycle completed by the van der Pol oscillator; a period that depends on the inner frequency $f_j$ of the oscillator, the amplitude $A$ and the frequency $f_0$ of the forcing function. The parameter $p$ controls the amplitude of the oscillations and $\mu$ controls the degree of nonlinearity of the oscillator. These latter two parameters can be assumed constant in the first approximation of the model to actual walking.

## 3.2    Parameter values

The statistics of the stochastic CPG are incorporated into the SCPG model by assuming the frequency $f_j$ of the van der Pol oscillator is given by

$$f_j = f_0 + \gamma X_j, \qquad (4)$$

where the values $X_j$ are generated by a random walk on a network of neurons. We suppose that the network of neurons is exponentially correlated and, for simplicity, we assume that it is given by a Markov chain, as done in Hausdorff et al. [11,24]. The nodes of this Markov chain are progressively activated by a random walker. The sequence $\{X_j\}$ are the values of the action potential at the nodes of the Markov chain progressively activated at the times $j = 1,2,..,N$. The exponential correlation length of the Markov chain is given by $r_0$. The autocorrelation function for the time sequence $\{X_j\}$ is not exponential but is calculated to be [7]

$$C_X(J) \propto \langle X_{j+J} X_j \rangle \propto \exp[Y/2] erfc[\sqrt{Y/2}] \qquad (5)$$

where the brackets denote an average over the random walk statistics, $erfc[x]$ is the complementary error function and $Y = J(\rho/r_0)^2$, where $J$ is the shift in time. The quantity $\rho$ is the mean length of a jump made by the random walker and, as we mentioned, $r_0$ is the correlation length of the network of neurons.  Fig. 3 shows that the autocorrelation function $C_X(J)$ converges to 1 with zero slope, for small $J$, and asymptotically converges to a inverse power law with Hölder exponent $h = -0.25$, for large $J$. The inverse power-law character of the correlation function (5) suggests the interpretation that asymptotically the series $\{X_j\}$, generated by the SCPG, is a fractal stochastic process.

The correlation length of the stochastic CPG can be changed to account for different modes of locomotion. The normal gait velocity, for a relaxed individual, with a mean frequency $f_0 = f_{0,n}$, has minimum correlation between the nodes of the CNS. An abnormal gait velocity, faster or slower than the normal one, with mean frequencies $f_{0,f}$ and $f_{0,s}$ respectively, increases the stress on the MCS with an increase of the correlation between the nodes of the CNS. We model this effect by assuming that $r_0$ is a function of gait frequency:

$$r_0 = r_{0,n} [ 1 + B (f_0 - f_{0,n})^2 ], \qquad (6)$$

where $r_{0,n}$ is the correlation length among the nodes of the CNS at normal gait velocity and $B$ is a positive constant that determines the degree of influence of frequency change on the correlation length. When the mean gait frequency $f_0$ is larger or smaller than the normal mean frequency $f_{0,n}$, the correlation length $r_0$ increases. The increase of the parameter $r_0$ with the gait frequency according to Eq. (6) leads to a decrease of the scaled variable $Y$ used in Eq. (5).

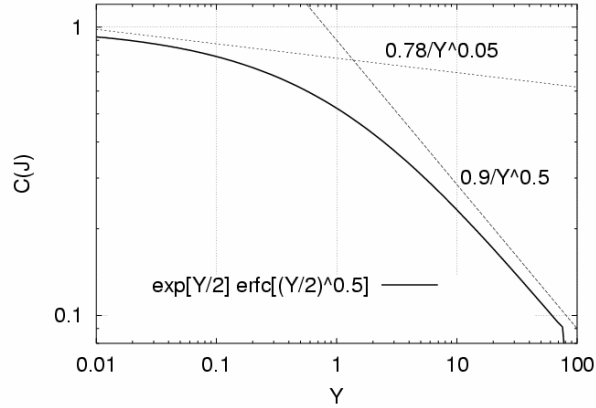

**Figure 3**: Autocorrelation function of the stochastic CPG, Eq. (5), with $Y = J (\rho/r_0)^2$.

## 4. Model Calculations

We determine the fractal exponents by using a fixed number of steps $J$ [5], consequently a decrease in $Y$ leads to a shift of the fitting range of the $J$ steps towards a region where the curve of the autocorrelation function Eq.(5) shown in Fig. 3, is characterized by a higher slope coefficient and a higher curvature. A higher slope coefficient may be interpreted as a higher fractal dimension and a higher curvature of the autocorrelation function may be interpreted as an increase of the multifractal properties of the signal. Therefore, we expect that the SCPG model predicts a slight increase of the Hölder exponents, as well as a slight increase of the multifractal properties, when the gait frequency deviates from normal. This change in the Hölder exponent mimics the behavior observed in the data. Moreover, a decrease of the parameter $r_{0,n}$ among the nodes of the central nervous system in the normal-relaxed condition implies an increase of the variable $Y$. More importantly, neurodegeneration may be interpreted as a decrease of the parameter $r_{0,n}$. According to the asymptotic behavior of the autocorrelation function $C_X(J)$ of Eq. (5), an increase of the variable $Y$ yields to a more monofractal sequence with a Hölder exponent $h = -0.25$, that is what we find on average for empirical data for the elderly or those with neurodegenerative diseases.

Figure 4 shows that the computer-generated stride interval time series produced by the SCPG model in different situations are similar to the phenomenological data shown in Fig 1. We assume in the calculation that the average periods of the normal, slow and fast gaits are 1.1, 1.45 and 0.95 seconds, respectively. These gait periods are compatible with the experimentally realized normal, slow and fast human gaits shown in Fig. 1. The other parameters used in the model are $\rho = 25$, $r_{0,n} = 25$, $B = 50$ and $\gamma = 0.02$ in the appropriate units. For the natural normal gait we use $A = 1$ and for both slower and faster gait, we use $A = 2$. For periodically triggered gait, we increase the intensity $A$ of the driver of the van der Pol oscillator to simulate the increase in stress on the MCS produced by the metronome. We use $A = 4$ for normal velocity and $A = 8$ for both slower and faster gait. The driver therefore models the conscious control of walking.

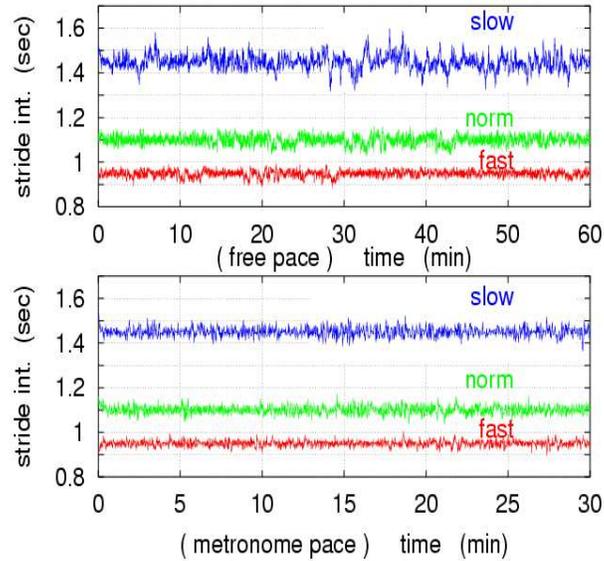

**Figure 4:** Typical computer-generated stride interval time series using the SCPG model in the freely walking and metronome paced conditions for normal, slow and fast pace.

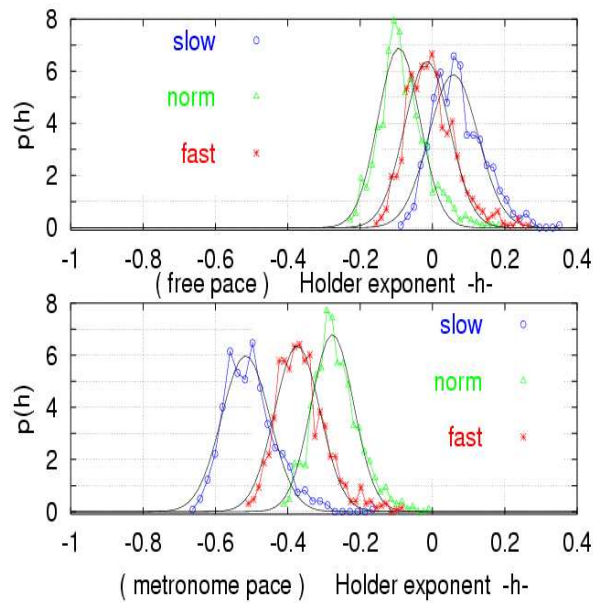

**Figure 5**: Typical Hölder exponent histograms for computer-generated stride interval series using the SCPG model in the freely walking and metronome-paced conditions for normal, slow and fast paces. The parameters of the SCPG model were chosen in such a way to approximately reproduce

the average behavior of the fractal and multifractal properties of the phenomenological data. The histograms are well-fitted with Gaussian functions.

Figure 5 shows histograms of distributions of the Hölder exponents for the computer-simulated gaits using the SCPG model depicted in Fig. 4. The parameters of the SCPG model were chosen in such a way that the model approximates the average behavior of the fractal and multifractal properties of the phenomenological data [5,14], *cf.* Fig 2. These properties were summarized above. Note the change in order along the exponent axis of the three modes of walking in the unconstrained and metronome synchronized cases, in both the data and the simulation.

## 5. Discussion

In summary, the stride interval of human gait presents a complex behavior that depends on many factors. Walking is a strongly correlated neuronal and biomechanical phenomenon which may be significantly influenced by two different stress mechanisms: (i) a natural stress that increases the correlation of the nervous system and regulates the motion at the changing of the gait regime from a normal relaxed condition, to a consciously forced slower or faster gait regime; (ii) a psychological stress due to the constraint of following a fixed external cadence such as a metronome. A psychological control, like that induced by a metronome, breaks the long-time correlation of the natural pace and generates a large fractal variability of the gait regime.

The SCPG model is able to mimic the complexity of the stride interval sequences of human gait under the several conditions of slow, normal and fast regimes for both walking freely and keeping the beat of a metronome. The model is based on the assumptions that human locomotion is regulated by both the CNS and by the INS. A network of neurons produces a fractal output that is correlated according to the level of physiologic stress and this network is coupled to the INS that generates the rhythmic process of the pace. The combination of the two systems, CNS and INS, controls walking and the variability of the gait cycle. It is the period of the gait cycle that is measured in the data sets and the SCPG model faithfully reproduces the stochastic and fractal characteristics of the phenomenological data. The correlation length in the SCPG determines the natural stress discussed in (a), whereas the amplitude of the driver models the psychological stress of the metronome in (b).

Finally, the SCPG correctly prognosticates that the decrease in average of the long-time correlation and of the multifractality of the stride interval time series for the elderly or for those with neurodegenerative diseases can be understood as a decrease of the correlation length among the neurons of the MCS due to neurodegeneration.